\documentclass[aps,prl,onecolumn,showpacs]{revtex4}

\usepackage{amsmath}
\usepackage{amssymb}
\usepackage{graphicx}% Include figure files
\usepackage{dcolumn}% Align table columns on decimal point
\usepackage{bm}% bold math

\newcommand{\prt}{\partial}

\begin{document}

\title{
Supersonic flow of a Bose-Einstein condensate past an oscillating attractive-repulsive obstacle }
\author{E.G. Khamis$^{1,2}$}
\email{e.g.khamis@lboro.ac.uk}
\author{A. Gammal$^{1}$}
\email{gammal@if.usp.br}

\affiliation{ 
$^1$ Instituto de F\'{\i}sica, Universidade de S\~{a}o Paulo, 05508-090, S\~{a}o Paulo, Brazil \\
$^2$ Department of Mathematical Sciences, Loughborough University, Loughborough, LE11 3TU, United Kingdom \\
}

\date{\today}

\begin{abstract} 
We investigate by numerical simulations the pattern formation after an 
oscillating attractive-repulsive obstacle inserted into the flow of a Bose-Einstein condensate. 
For slow oscillations we observe a complex emission of vortex dipoles. For moderate 
oscillations organized lined up vortex dipoles are emitted. For high frequencies no dipoles are 
observed but only lined up dark fragments. The results shows that the drag force turns 
negative for sufficiently high frequency. We also successfully model the {\it ship waves} in 
front of the obstacle. In the limit of very fast oscillations all the excitations of the 
system tend to vanish. 
\end{abstract}


\maketitle

{\bf I. INTRODUCTION} 
\bigskip

The realization of Bose-Einstein condensate (BEC) in atomic gases have boosted 
intense theoretical and experimental investigation about its exotic 
properties. BEC is a paradigm of a quantum fluid and in weak interacting case can 
well be described by the Gross-Pitaesvkii equation \cite{dalfovo}. Eventually 
BEC spread to other systems like exciton-polaritons \cite{exciton2006}, offering 
new possibilities for experimental tests. An interesting feature of a quantum 
fluid is its contrasting behavior as compared with a classical fluid.  The 
flow of a quantum fluid past an obstacle only generates drag force above a 
certain subsonic critical velocity and the energy can be dissipated into 
collective excitations of the fluid. This dissipation can be inferred from 
numerical experiments by the mean drag on the obstacle~\cite{frisch}. A 
superfluid behavior is revealed below this velocity where nucleation of vortices 
never occur and no excitations are generated 
\cite{jackson,winiecki,nore,stiessberger,huepe,aftalion}. For an appropriate 
velocity and size of the obstacle, a B\'{e}nard-von K\'{a}rm\'{a}m vortex street 
can be generated~\cite{sasaki}.

There is also a supersonic critical velocity where oblique vortex streets are 
transformed into stable oblique dark solitons \cite{kp-08}. For higher 
velocities, the general picture of the diffraction pattern in the supersonic 
flow past a disk-shaped impenetrable obstacle consists of two different parts 
separated by the Mach (or Cherenkov) cone~\cite{egk-prl06}. Outside the Mach 
cone there is a region of linear waves that we will refer them simply as {\it 
ship waves} \cite{carusotto06,shipwaves}. Inside the Mach cone a pair of oblique 
dark solitons is gradually formed behind the obstacle if the radius of the 
obstacle is of healing length order. For greater radius more pairs of oblique 
solitons are generated. Interaction of solitons was studied in 
\cite{annibale2011,khamis2012} where it was found that the angle between dark 
solitons decreases as the obstacle radius increases for a fixed supersonic 
velocity of the flow. In previous experimental works 
\cite{carusotto06,amo-2009,amo-2009-2} the existence of such nonlinear 
structures were suggested. However, only recently the generation of stable 
oblique dark solitons was experimentally demonstrated in the flow of a 
Bose-Einstein condensate of exciton-polaritons past an obstacle 
\cite{amo-2011,grosso2011}. A numerical study to support these experimental 
findings was done in \cite{pigeon}, and the observation of vortex dipoles in an 
oblate atomic Bose-Einstein condensate \cite{brian} suggests that the supersonic 
studies can also be carried in this system.

In atomic BEC, obstacles are typically represented by detuned lasers that can be 
effectively be attractive (red-detuned) or repulsive (blue-detuned) obstacles, 
by the use of Feshbach resonances.  The first numerical study of attractive 
obstacles was done in Ref. \cite{jacksonattractive} where it was established the 
critical velocity to the formation of vortices and corrects the velocity found 
in \cite{frisch} in the case of repulsive obstacles. Numerical studies revealed 
that turbulence can also be achieved and studied by spatial oscillation of a 
repulsive obstacle \cite{tsubota}. A clever way to control the formation of 
vortices moving attractive and repulsive laser beams was proposed in 
\cite{saito}. There is a special form of moving potential that no radiation is 
generated at supersonic velocities \cite{law}. The disappearance of gray soliton 
and phonon excitations was demonstrated in \cite{radouani} by oscillating a 
repulsive obstacle in a quasi-1D trapped BEC at high obstacle velocities. It was 
found in \cite{saito2012} that vibration of an obstacle modulates the vortex 
street, theoretically predicted in~\cite{sasaki}, breaking a symmetry.

In the present work, we study the flow of a BEC past an oscillating attractive 
and repulsive obstacle. The motivation is to answer the question {\it can we get 
rid of drag for very fast oscillations?}  We investigate different regimes from 
slow to very fast oscillations. Since we are working in the supersonic regime we 
can divide the study inside and outside the Mach cone as follows.

\bigskip
{\bf II. MODEL EQUATIONS} 
\bigskip

We consider the flow of an atomic  Bose-Einstein condensate (BEC) past 
an obstacle in the framework of Gross-Pitaevskii (GP) 
mean field approach. In the rest frame, the condensate is well 
described by the macroscopic wave function $\Psi\equiv\Psi(x,y,z,t)$ 
obeying the time-dependent GP equation 
\begin{equation}\label{dimension3}
i \hbar \frac{\prt\Psi}{\prt t}=-\frac{\hbar^2}{2m}\nabla^2\Psi
+U_{ext}\,\Psi+\frac{4\pi a\hbar^2}{m}|\Psi|^2\Psi,
\end{equation}
where $\nabla^2 \equiv \partial^2_x+\partial^2_y+\partial^2_z$, the external 
potential $U_{ext}=U_{trap}(x,y,z)+U(x+vt,y,z,\Omega t)$ is represented 
by the sum of a harmonic trap $U_{trap}$ and a time-dependent obstacle 
potential $U$ that oscillates with frequency $\Omega$, 
$m$ is the atomic mass and $a$ is the s-wave scattering length. 

We will limit our study to the case of the quasi-2D limit, i.e., we have a strong harmonic confinement 
in the $z$ direction. In this regime we can approximate 
$\Psi(x,y,z,t)=\psi(x,y,t)\phi(z)e^{-i\mu_z/\hbar}$, where $\phi(z)$ and $\mu_z$ are the ground state 
and energy respectively for the confinement in $z$ direction \cite{perez,saito}. Substituting 
in Eq.(\ref{dimension3}) and integrating in $z$ direction we obtain
\begin{equation}\label{dimension2}
i \hbar \frac{\prt\psi}{\prt t}=-\frac{\hbar^2}{2m}(\prt^2_x+\prt^2_y)\psi
+U\, \psi+g|\psi|^2\psi,
\end{equation}
where $g=4\pi a\hbar^2m^{-1}\int\phi^4(z)dz$ is the effective interaction in two-dimensions. 
We consider here that the obstacle runs close to the center of the trap. In this region the condensate 
is almost homogeneous and the potential in $x$ and $y$ directions is weak as compared to 
the obstacle potential. So the harmonic potential is neglected for studying the excitation caused by 
the obstacle.  

We introduce dimensionless variables ${\tilde x}=x/\xi$,
${\tilde y}=y/\xi$, ${\tilde t}=gn_0t/\hbar$, ${\tilde \psi}=\psi/\sqrt{n_0}$, 
${\tilde U}=U/gn_0$, ${\tilde \Omega}=\Omega \hbar/gn_0$, the
Mach velocity $M=v/c_s$, where $n_0$ is a characteristic 2D density of atoms at the center of the trap,  
$\xi=\hbar/\sqrt{mn_0g}$ is the characteristic length and the sound velocity 
$c_s=\hbar/m\xi$. Typical experimental values are $\xi\sim0.3\mu$m and $\hbar/gn_0\sim0.18$ms 
\cite{brian}.
Thus for ${\tilde \Omega\sim 1}$ we have oscillations of the order of kHz well within experimental 
reach. 
Substituting in Eq.~(\ref{dimension2}) and after dropping the tildes for convenience we get
\begin{equation}\label{4}
    i\frac{\prt\psi}{\prt t}=-\frac{1}{2}(\prt^2_x+\prt^2_y)\psi
+U\psi +|\psi|^2\psi\, , 
\end{equation} 
where $U=U(x+Mt,y,\Omega t)$. The energy is given by 
\begin{equation}
E(t)=\int dxdy \left[ {-\frac{1}{2}|\nabla_{x,y} \psi|^2+U|\psi|^2+\frac12|\psi|^4} \right]\, ,
\end{equation}
and the rate of energy in time is given by 
\begin{equation}
\frac{d E}{d t}=M\int dx dy |\psi|^2 \frac{\prt U(\overline{x},y,\Omega t)}{\prt \overline{x}} |_{\overline{x}=x+Mt}
                      +\int dx dy |\psi|^2   \frac{\prt U(\overline{x},y,\Omega t)}{\prt t} |_{\overline{x}=x+Mt} \, .  
\end{equation}
Turning off the oscillation ($\Omega=0$) the second term vanishes and we identify the first term 
as $M$ times the drag force.
Also when $M=0$ only the second term is responsible for the excitation of the system.

For computational purposes, in Eq.~(\ref{4}) we make a global phase transformation $\psi'=e^{it}\psi$ 
and later a Galilean transformation $x'=x+Mt$, $t'=t$ leading to
\begin{equation}\label{7}
i\frac{\prt\psi}{\prt t}= -\frac{1}{2}(\prt^2_x+\prt^2_y)\psi
-iM\prt_x\psi-\psi+|\psi|^2\psi +U \psi\,,  
\end{equation} 
where $U=U(x,y,\Omega t)$, the primes were omitted for convenience 
and subscripts here means derivatives. This equation describes the system 
in the obstacle reference frame.  

The obstacle is a laser beam that continuously oscillates from 
blue-detuned to red-detuned and vice-versa,  which can be written as
\begin{equation}
U(x,y,\Omega t)=U_0 \cos(\Omega t) \exp \left[ \frac{-2\left(x^2+y^2\right)}{w_0^2} \right],
\end{equation}
where $U_0$ and $w_0$ are the amplitude and the beam waist of the 
laser, respectively, and $\Omega=2\pi/T$ is the oscillation frequency of the 
detuning in a period $T$. 

\bigskip
{\bf III. INSIDE THE MACH CONE}  
\bigskip

For $\Omega=0$ Eq.~(\ref{7}) supports stable oblique 
solitons \cite{egk-prl06,kp-08,khamis2012} in the obstacle frame as
\begin{equation} \psi(x,y)=\frac{1+e^{\zeta+2i\alpha}}{1+e^\zeta} \,, \label{exactss}
\end{equation}
where $e^{i\alpha}\equiv\lambda+i\nu$, $\zeta\equiv 2\nu[x\sin\theta-y\cos\theta]$, $\nu\equiv\sqrt{1-\lambda^2}$,
$\lambda\equiv M\sin\theta$, and $\theta$ is the angle between the soliton and the horizontal axis.
$M\sin\theta=\pm1$ defines the Mach cone and thus solitons can be found only in the region
$-\arcsin(1/M)<\theta<\arcsin(1/M)$.

\begin{figure}[ht]
\includegraphics[width=5cm,clip]{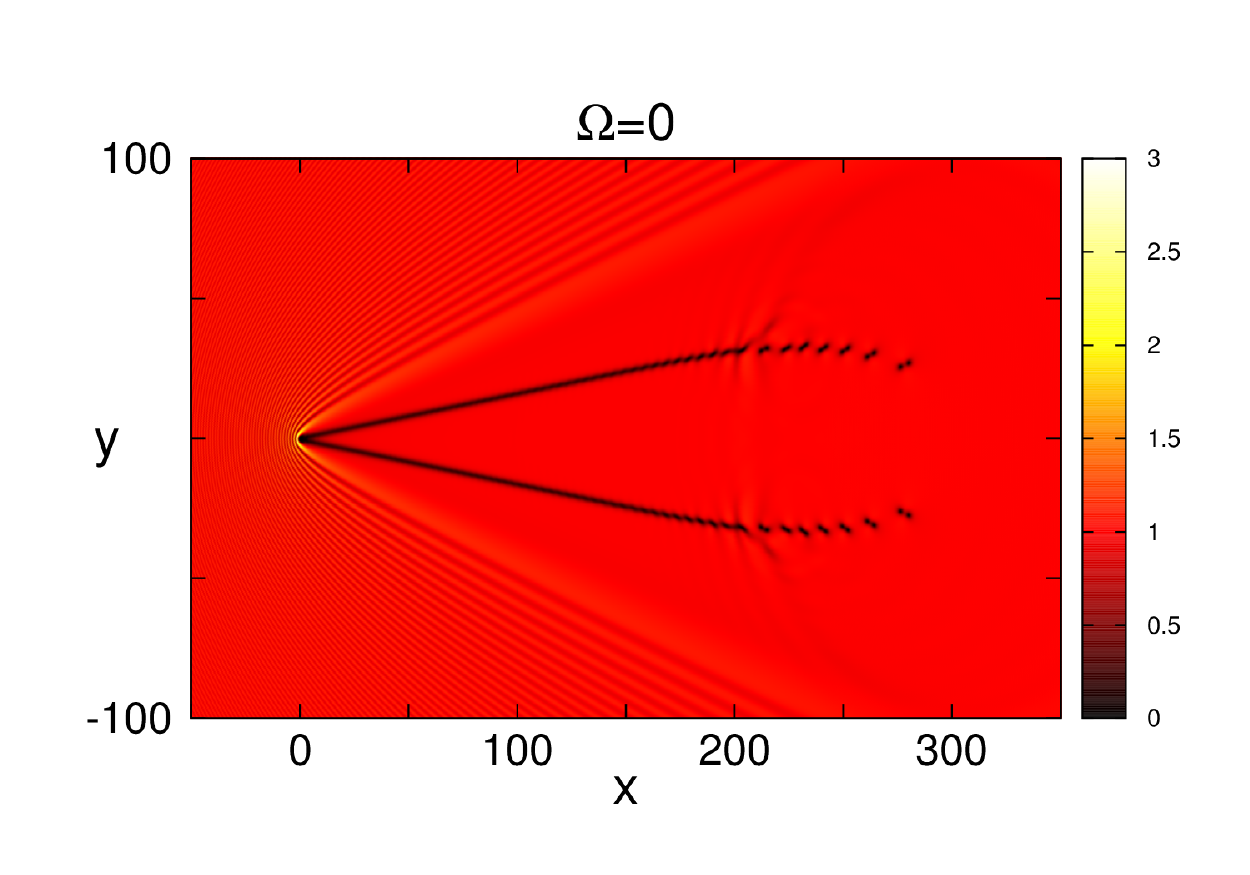}
\includegraphics[width=5cm,clip]{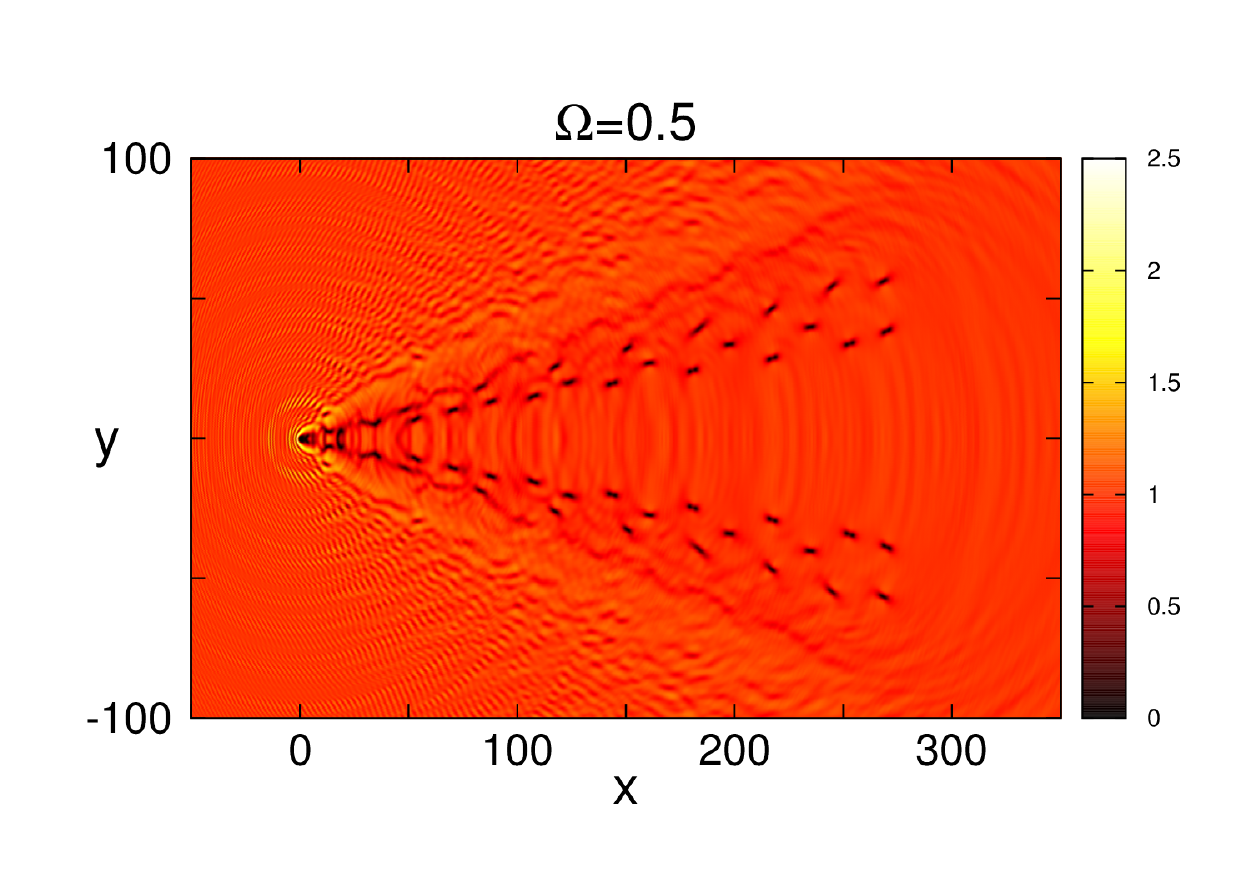}
\includegraphics[width=5cm,clip]{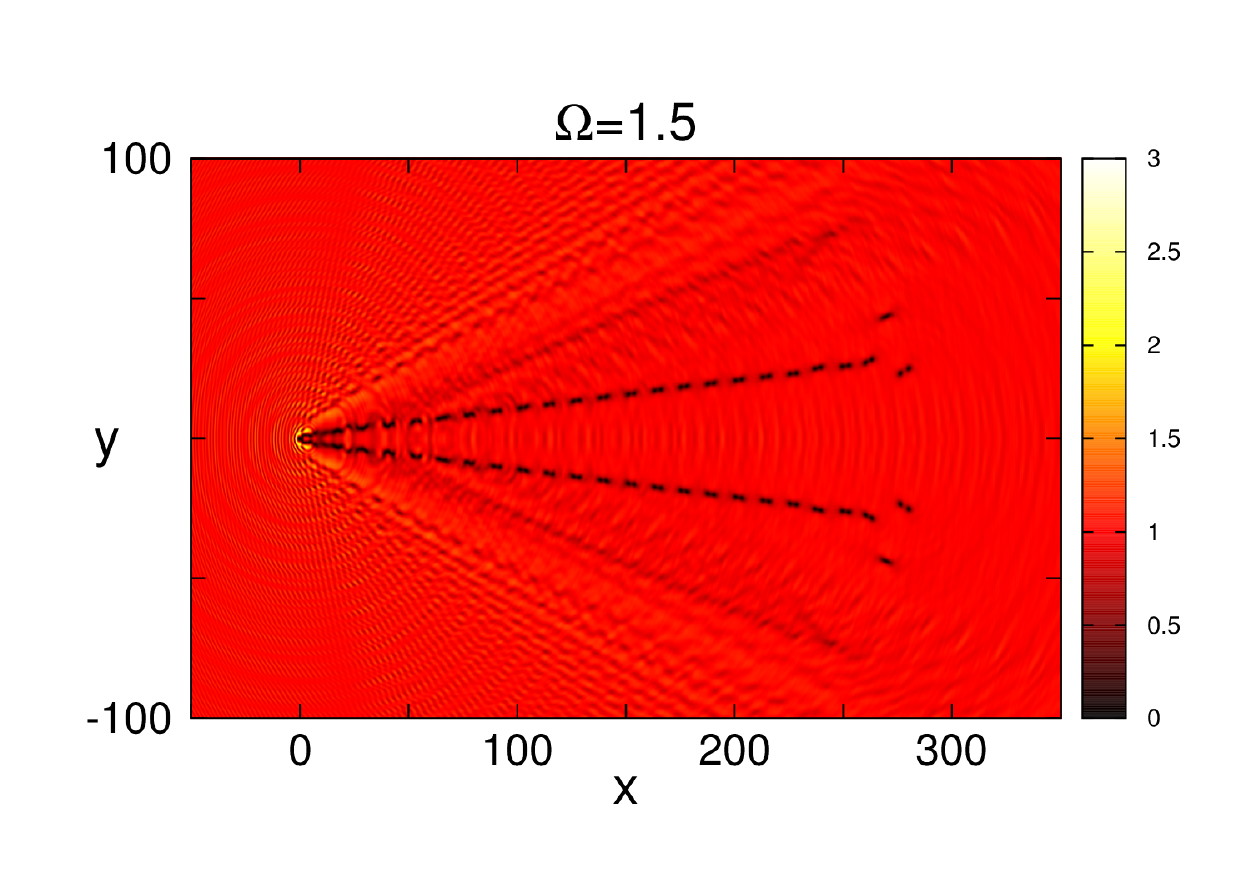}
\includegraphics[width=5cm,clip]{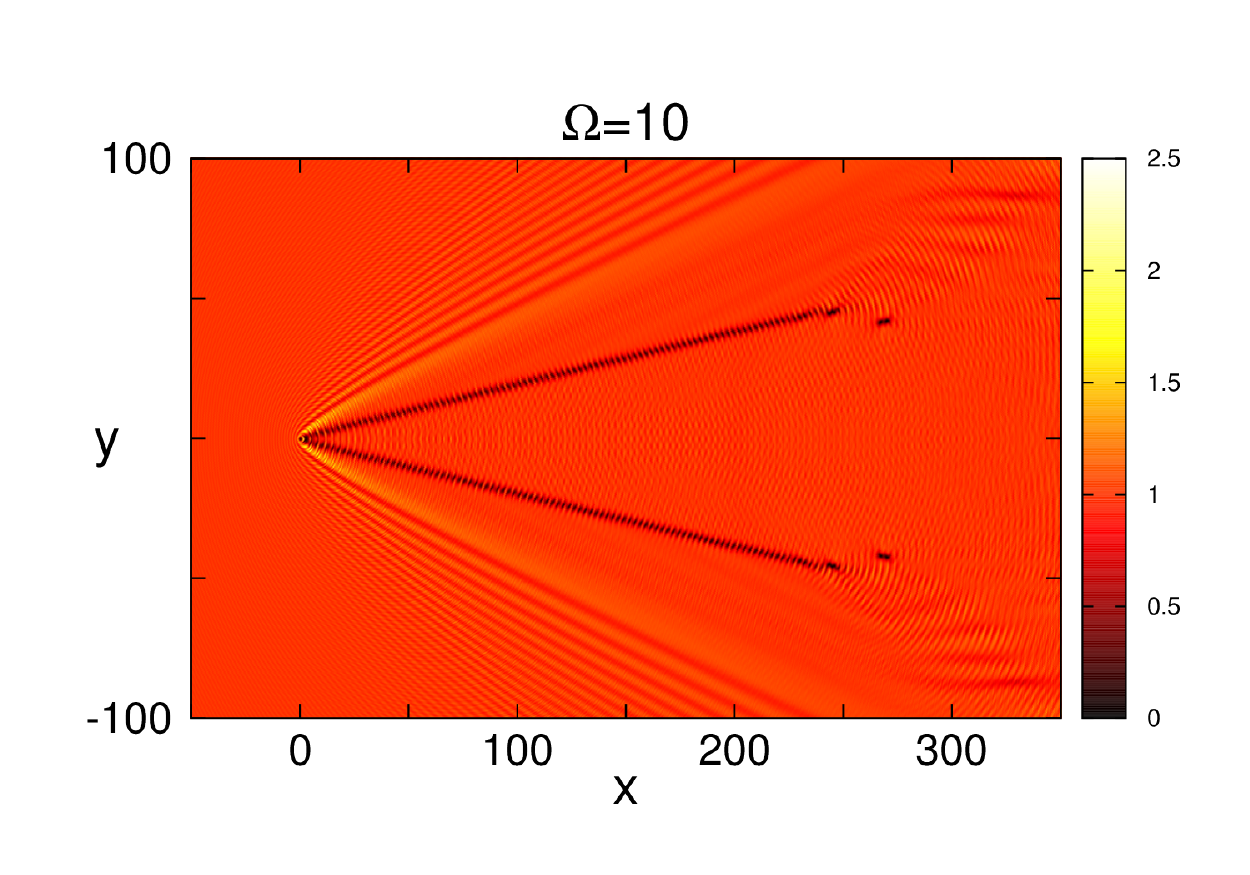}
\includegraphics[width=5cm,clip]{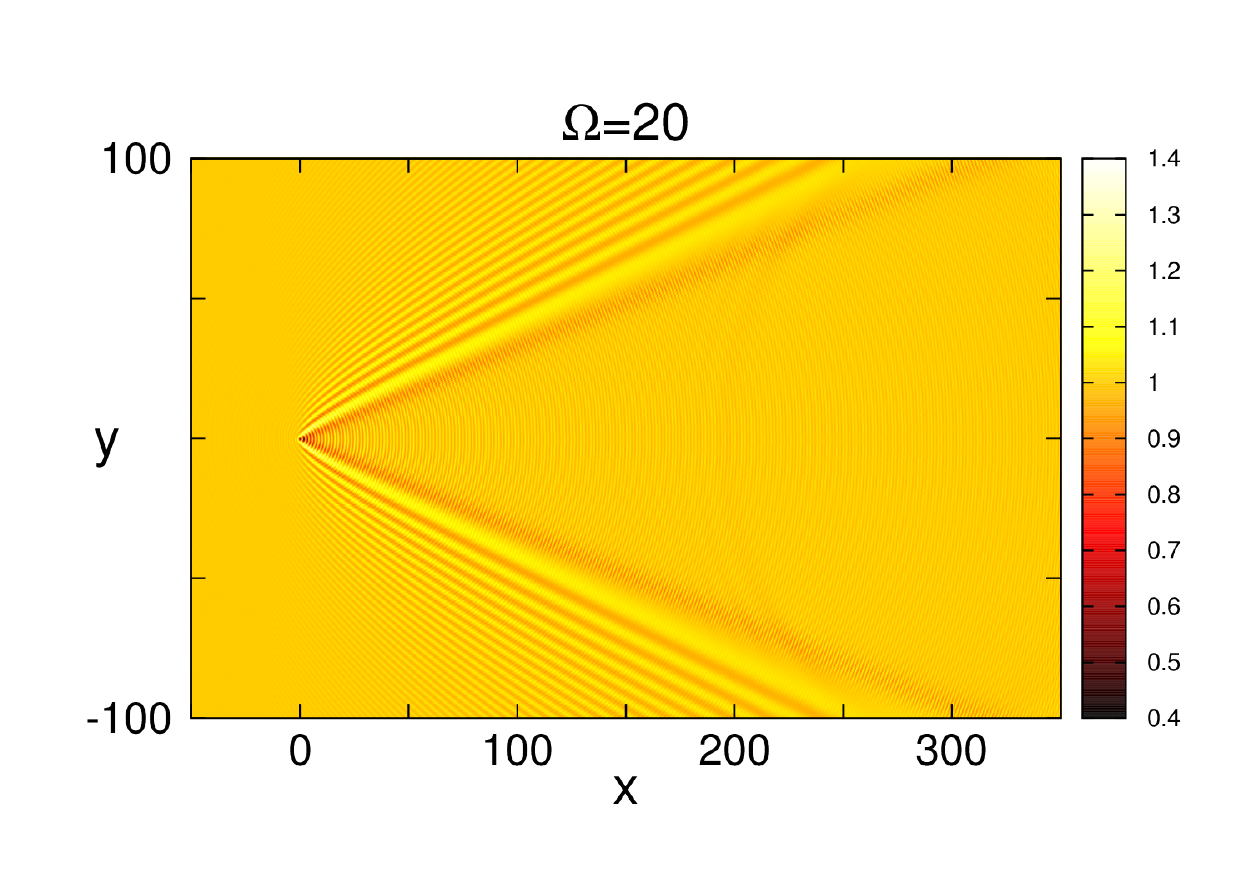}
\caption{Pictures of the diffraction pattern at fixed time $t=100$ 
with the set of parameters: Mach number $M=3$, potential intensity $U_0=25$ 
(the laser beam starts repulsive) and width $w_0=1$ for different frequencies 
$\Omega=0$, 0.5, 1.5, 10 and 20.
}
\label{fig3}
\end{figure}

\begin{figure}[ht]
\includegraphics[width=5cm,clip]{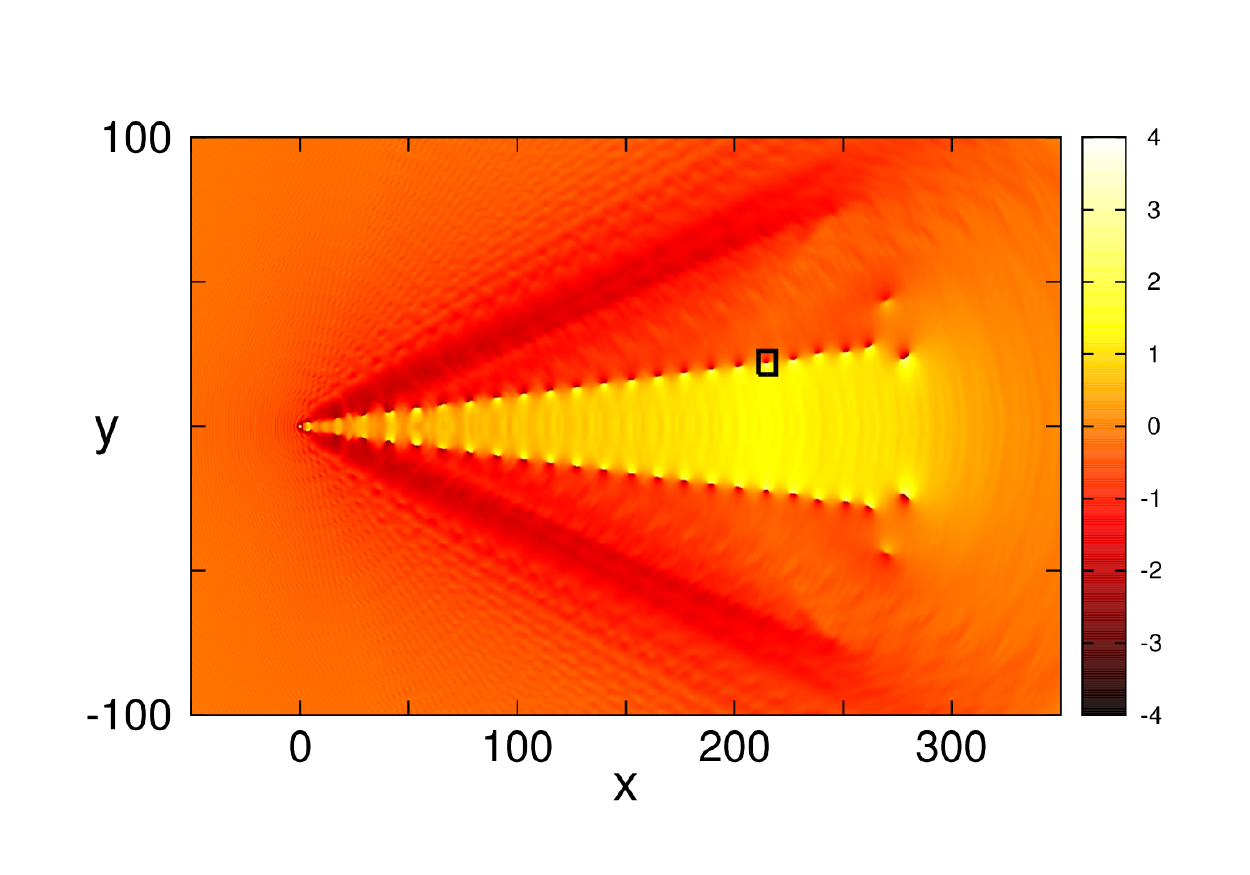}
\includegraphics[width=5cm,clip]{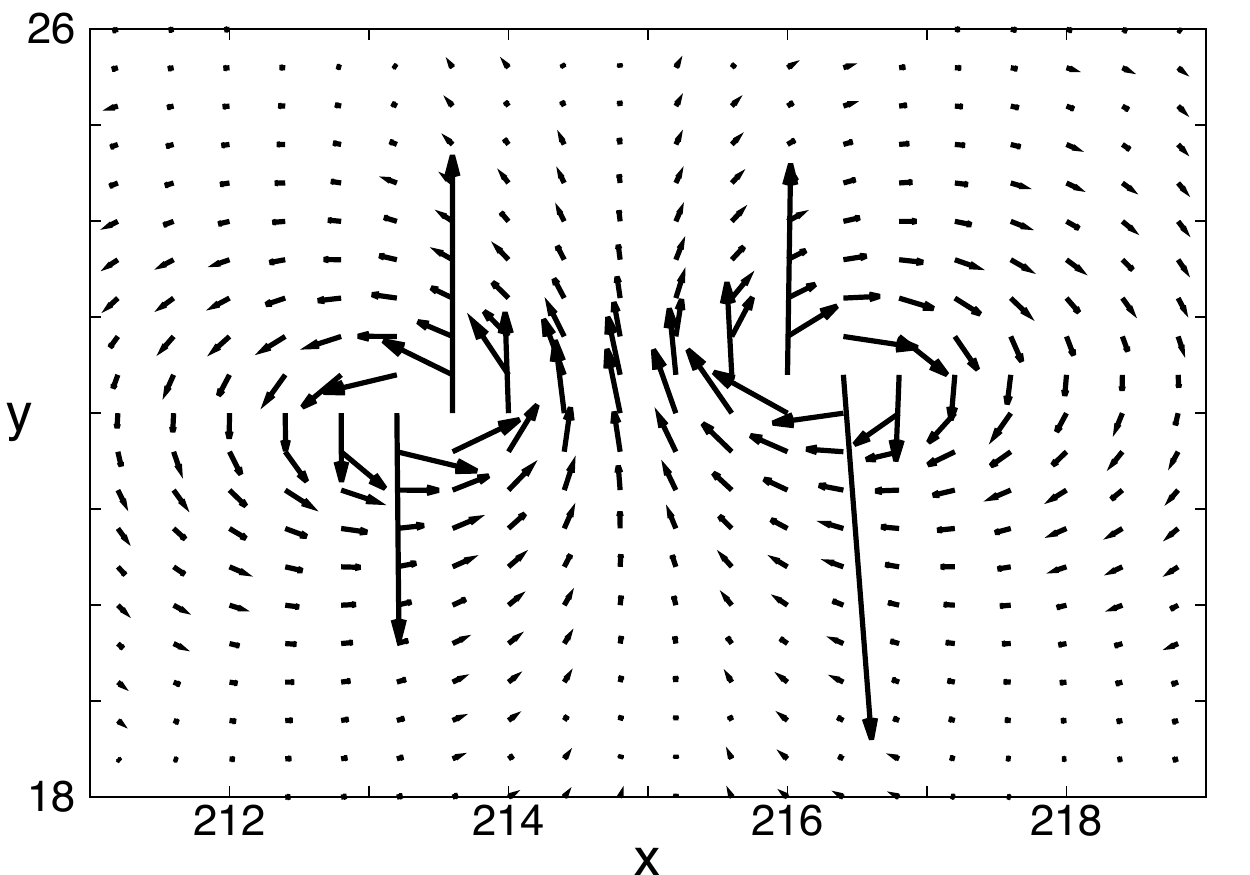}
\caption{Left panel: Phase of the diffraction pattern that corresponds to Fig.~\ref{fig3} 
for frequency $\Omega=1.5$ which provides us evidence of vortex dipole formation. 
Right panel: inset showing vector velocities of the vortex dipole in the selected region. 
Note that the velocity in $x$ direction was not shifted by the constant flow 
velocity, $M$, for visualization purposes.
}
\label{phase}
\end{figure}

We have solved the Eq.~(\ref{4}) numerically in the supersonic regime using $U_0=25$ and $w_0=1$. 
In Fig.~\ref{fig3} we show the results for supersonic flow for different oscillation frequencies. 
Fig.~\ref{fig3} depicts the case of $\Omega=0$ where we reproduce the formation of oblique dark 
solitons~\cite{egk-prl06}. Outside the Mach cone there is a stationary wave pattern created by 
interference of linear waves. Inside the Mach cone there are two oblique dark solitons that decay 
at the end points into vortices, situated symmetrically with respect to the direction of the flow. 
As we turn on the oscillations observe the emission of dark fragments. For $\Omega=0.5$ these 
fragments can be identified as vortex dipoles and form a pattern of ``5 in a dice". As the 
frequency is increased to $\Omega=1.5$ these fragments stand well aligned as vortex dipoles as can 
be identified by the phase plot (see Fig.~\ref{phase}). These dipoles are followed by a secondary 
radiation emission, identified as a straight line almost parallel to the Mach cone. As the 
frequency is further increased to $\Omega=10$ the fragments can no longer be identified neither as 
a single vortex nor as vortex dipoles. By looking at phase the fragments are identified as short 
gray solitons that propagate obliquely to the flow, analogous to the ones observed in 
Ref.~\cite{flayac2012}. To check the (non)vorticity character after some time of fragments 
formation we turned off the intensity of the obstacle and no decay into vortices were observed.

One can explain the general behavior as following. For $\Omega=1.5$ the oscillation 
acts as a ``chopper'' that turns {\it on} and {\it off} the dipole emission. In this 
specific case the {\it on} time is more than enough to generate vortex dipoles and thus 
we have excess of energy that is ejected as secondary radiation. In the fast oscillating 
regime ($\Omega=10$) the time the oscillation is {\it on} is not enough to form dipoles 
and just small dark solitons can be seen. As the frequency is around $\Omega=20$ 
practically no more fragments can be seen.  To check the consistency of our analysis we 
studied the number of fragments as a function of $\Omega$. One can estimate that rate 
of fragments emission is close to 1, meaning that at each period one fragment is emitted. 
The linear behavior confirms the modeling of the oscillating obstacle as a ``chopper''.

Thus we identified four regimes, for $\Omega\sim0.5$ we have strange patterns as 
the ``5 in a dice'', for $\Omega\sim1.5$ we have lined vortex-dipoles with secondary 
radiation, for $\Omega\sim10$ the dipoles are suppressed and give place to small dark 
solitons that propagate obliquely to the flow, finally for $\Omega\sim20$ practically no 
excitation can be seen inside the Mach cone. If the laser beam starts red-detuned 
(attractive) we have similar results. 

Next we develop the analytical theory to describe the remaining ship waves 
located outside the Mach cone and compare it with numerical 
simulations.

\bigskip 
{\bf IV. OUTSIDE THE MACH CONE} 
\bigskip 

Ship waves are formed in front of the obstacle. The theory for a 
non-oscillating obstacle was previously studied for a $\delta$ function  
in ref.\cite{khamis08}, where it was found that the density changes are given by 
\begin{equation}
\delta n=V_0 \, q(k,r,M) \times \cos \left( kr\cos\mu-\frac{\pi}{4} \right) 
%,\quad y=0, \quad x<0,
\end{equation}
with
\begin{equation}q(k,r,M)\equiv
\sqrt{\frac{2k}{\pi r}} \frac{ [(M^2-2)k^2+4(M^2-1)]^{1/4} }
     { [(M^2-2)k^2+6(M^2-1)]^{1/2} } \,,
\end{equation}
and 
\begin{equation}\label{old}
k\equiv 2\sqrt{M^2\cos^2{\eta}-1}.
\end{equation}

\begin{figure}[ht]
\includegraphics[width=3.5cm]{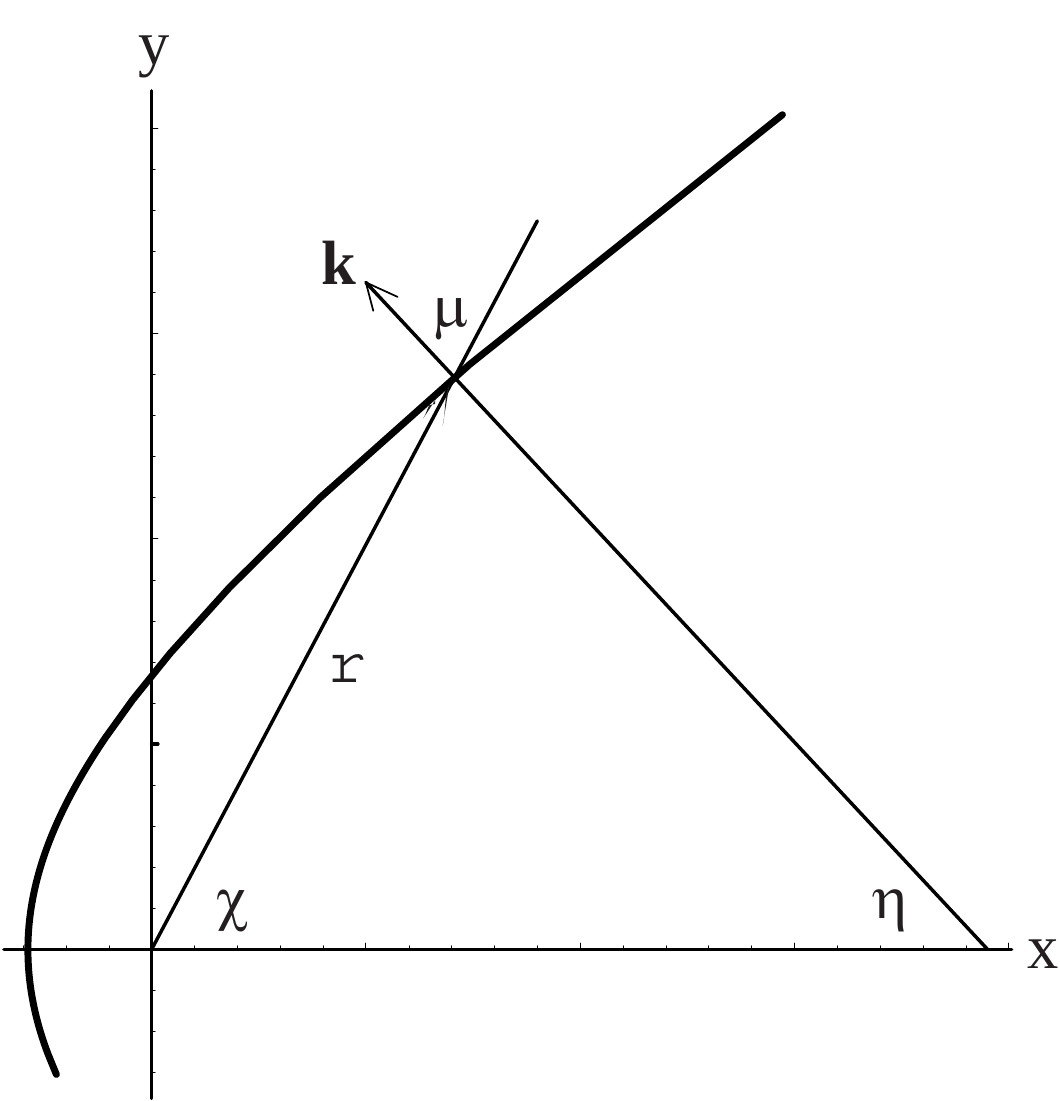}
\includegraphics[width=5cm]{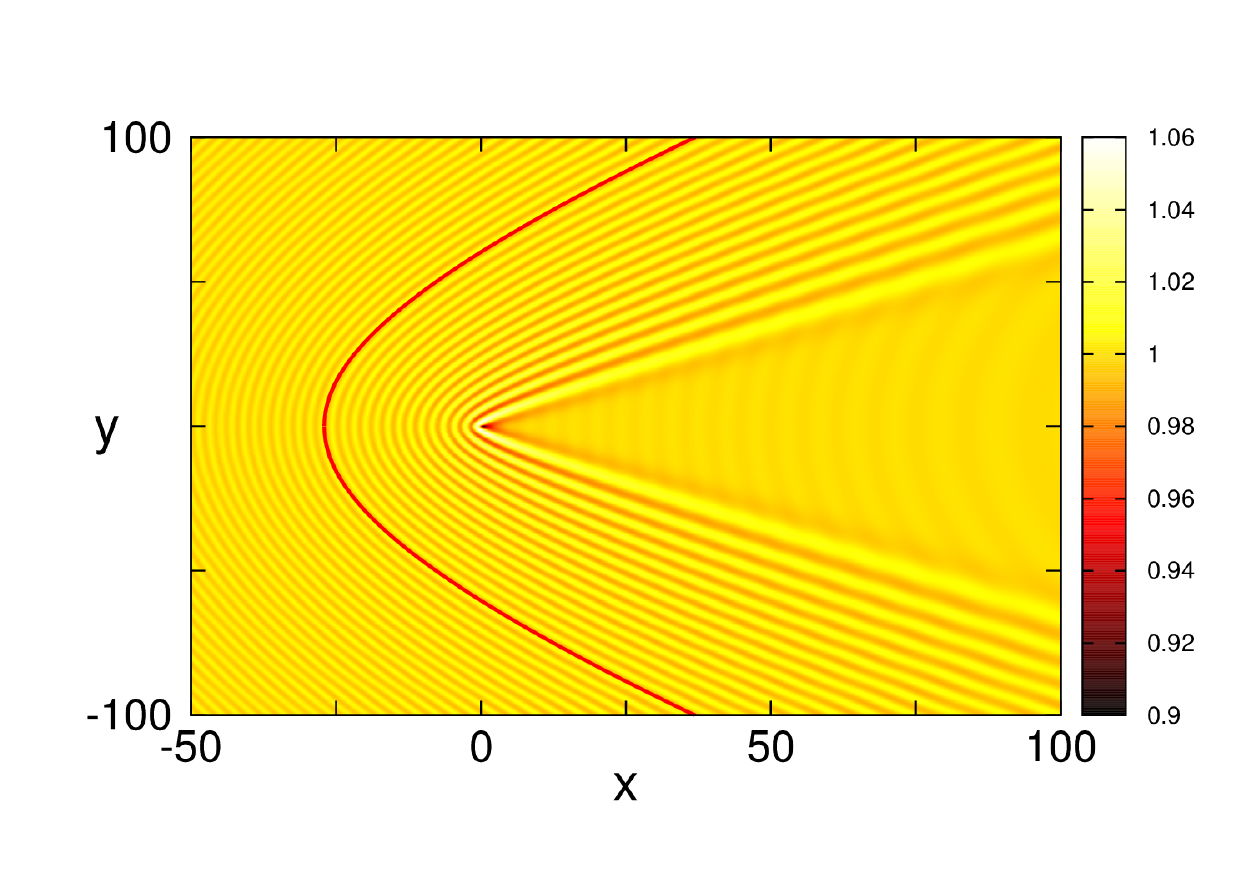}
\caption{Left panel: Coordinates that defines the radius-vector ${\bf r}$ and the wave 
vector ${\bf k}$. The latter one is normal to the wave front which is shown 
schematically by a curved line. 
Right panel: Numerically calculated wave pattern for a fast oscillating obstacle at 
fixed time $t=100$ with the set of parameters: $M=2$, $U_0=25$, $w_0=1$ and  
$\Omega=40$. Solid line (red) corresponds to linear analytical theory 
(Eq.~\ref{old2}) for the line of constant phase.
}
\label{fig1}
\end{figure}

The angles $\mu$ and $\eta$ are defined according to Fig.~\ref{fig1}, and 
Eq.~(\ref{old}) is valid if 
\begin{equation}
    -\arccos\left(\frac{1}{M}\right)\leq\eta\leq\arccos\left(\frac{1}{M}\right)~,
\end{equation}
so that the linear waves exist only in the region outside the Mach cone.

According to \cite{khamis08}, one can find the shape of the lines of 
constant phase (wave crests) $\Phi=kr\cos\mu$ in a parametric form 
\begin{equation}\label{old2}
\begin{split}
&x=r\cos\chi=\frac{4\Phi}{k^3}\cos\eta(1-M^2\cos2\eta),\\
&y=r\sin\chi=\frac{4\Phi}{k^3}\sin\eta(2M^2\cos^2\eta-1),
\end{split}
\end{equation}
so that for small values of $\eta$, corresponding to waves in front of the 
obstacle, these lines take a parabolic form 
\begin{equation}
 x(y)\cong -\frac{\Phi}{2\sqrt{M^2-1}}+\frac{(M^2-1)^{3/2}}{(2M^2-1)\Phi}y^2.
\end{equation}
Predictions of the analytical theory are compared with the numerically 
calculated wave pattern in Fig.~\ref{fig1} and excellent agreement is found. 
So, the theory previously developed in \cite{khamis08} remains valid even 
for a fast oscillating obstacle. 

For a fast oscillating we assume that the resulting ship 
waves can be computed by the Huygens principle, i.e., by the superposition of 
stationary densities generated by obstacles at different positions along the flow. 
Averaging over a period this can be 
expressed as 
\begin{equation} \delta n_{osc}=V_0 \, 
q(k,r,M)
\times \frac{1}{T}\int_0^T\cos(\Omega t) 
\cos \left( kr\cos\mu-\frac{\pi}{4}+kMt \right)~dt, 
\end{equation} 
where the term $kMt$ was added 
representing phase change due to the obstacle movement along the time.
After integration in time one obtains 
\begin{equation} \delta n_{osc}=\frac{1}{2\pi}V_0 \, q(k,r,M) 
\left( \frac{-\Omega kM}{\Omega^2-k^2M^2} \right)
\left[ \sin \left( kr\cos\mu-\frac{\pi}{4}+kMT \right) 
      -\sin \left( kr\cos\mu-\frac{\pi}{4} \right) \right]. 
\end{equation} 

\begin{figure}[ht]
\includegraphics[width=5cm,clip]{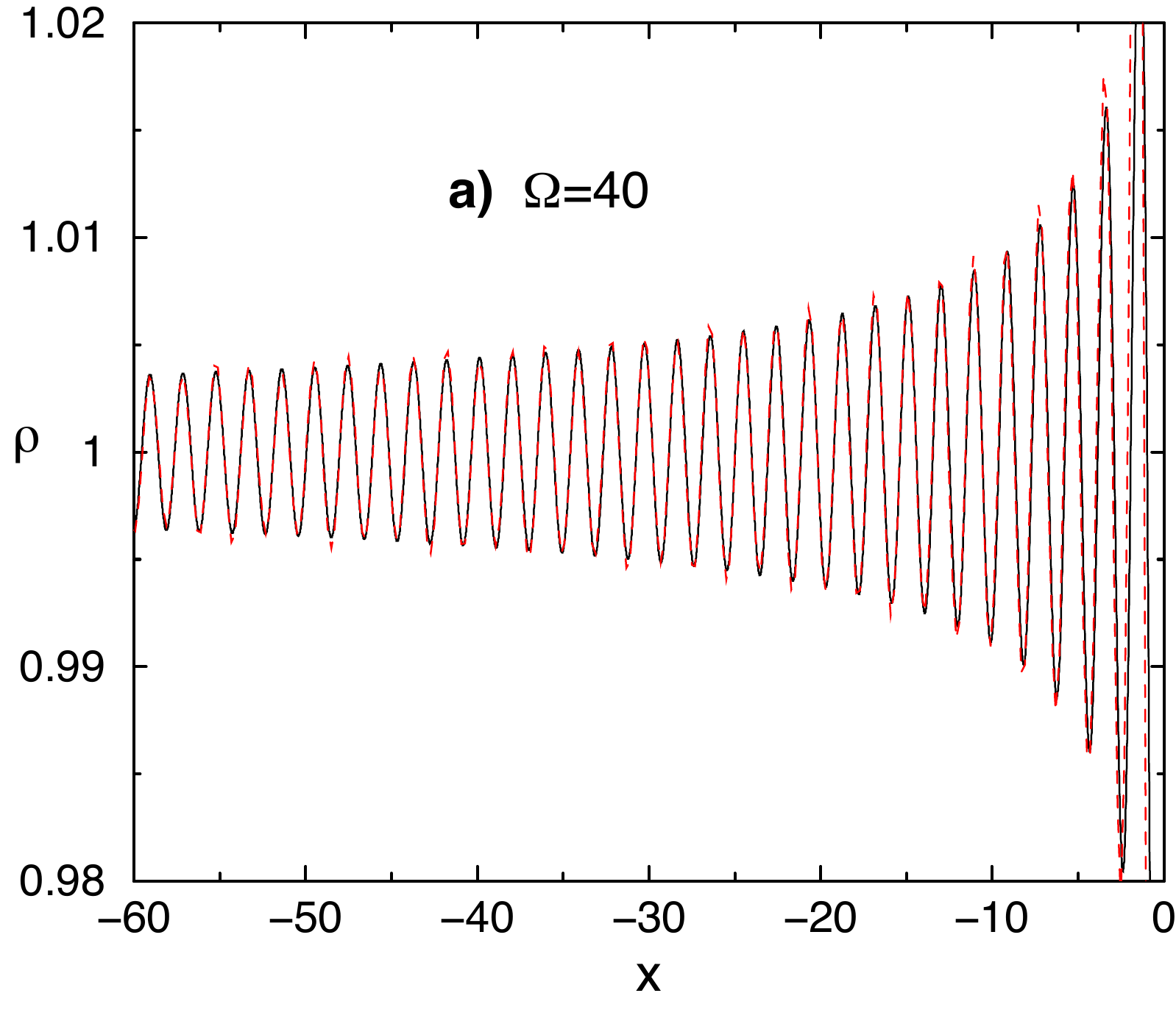}
\includegraphics[width=5cm,clip]{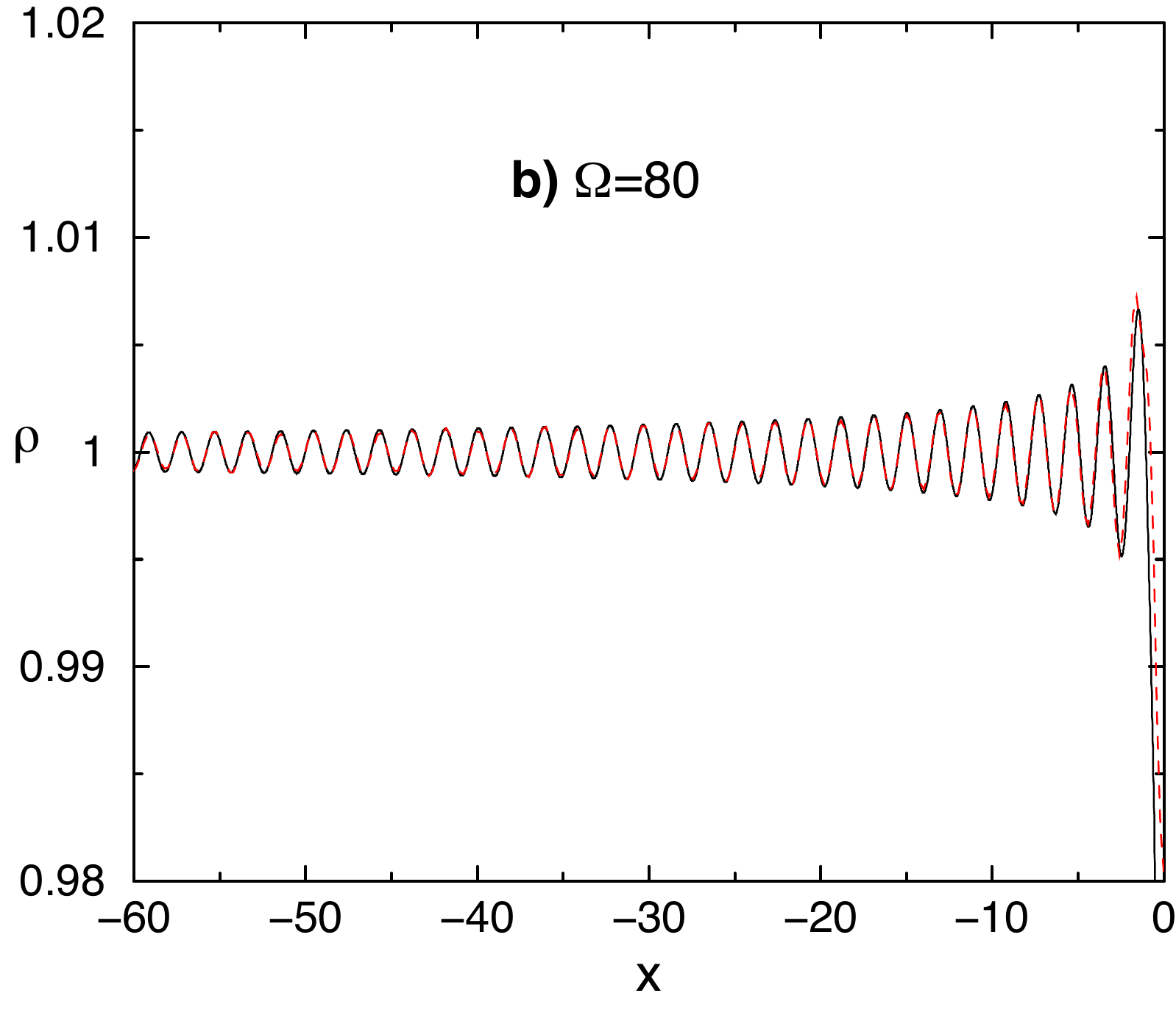}
\includegraphics[width=5.1cm,clip]{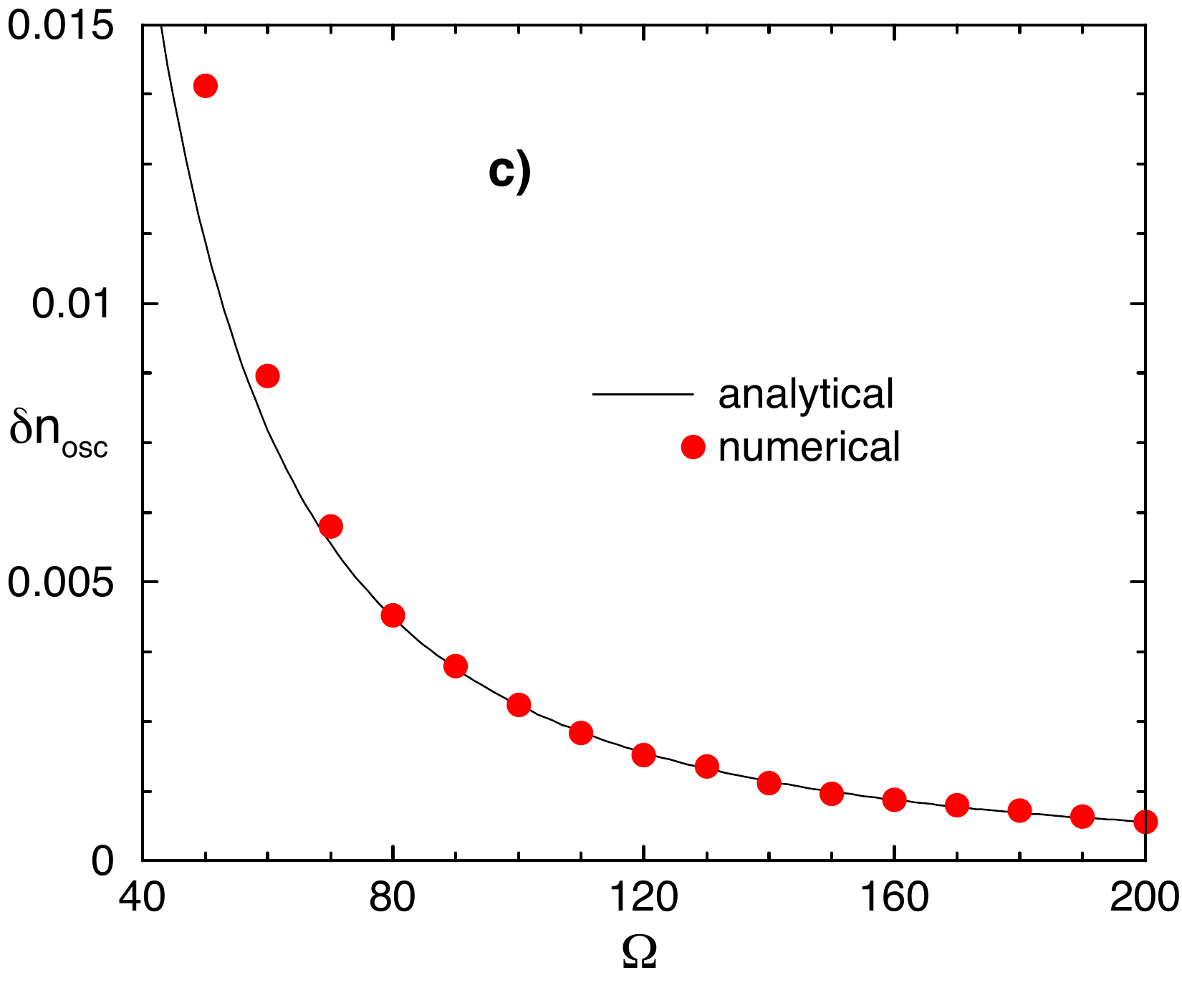}
\caption{
(a) and (b) Profile of intensity in front of the obstacle for $x<0$, 
$y=0$ with the set of parameters: $M=2$, $V_0=2$, $U_0=25$ 
(the laser beam starts repulsive) and $w_0=1$. 
Solid lines (black) correspond to linear analytical theory, Eq.~(\ref{ship}), 
and dashed lines (red) to numerical solution of Eq.~(\ref{4}). 
(c) The solid line corresponds to the higher magnitude value 
of the  $\delta n_{osc}$ close to $x=-40$ and the red circles 
correspond to this magnitude calculated numerically at the same position. 
}
\label{fig2}
\end{figure}

In the region in front of the obstacle where $y=0$ (i.e. $\eta=0$) 
and $x<0$, the 
wavelength $\lambda=2\pi/k_0$ is constant with $k_0=2\sqrt{M^2-1}$. 
Therefore $x(0)= -r\cos\mu$ and the perturbations of the condensate density 
take the form 
\begin{equation}\label{ship}
\delta n_{osc}=\frac{1}{2\pi}V_0 \,
q(k_0,r,M) 
\left( \frac{-\Omega k_0M}{\Omega^2-k_0^2M^2} \right)
\left[ \sin \left(-k_0x-\frac{\pi}{4}+\frac{2\pi k_0M}{\Omega}\right) 
      -\sin \left(-k_0x-\frac{\pi}{4}\right) \right], 
\end{equation}
where $r=|x|$.
The above formulae shows that increasing $\Omega$ the magnitude of the 
$\delta n_{osc}$ decreases as $1/\Omega$ and in the limit 
of $\Omega\rightarrow\infty$ the ship waves vanish. 
The plot illustrating this behavior is shown in Fig.~\ref{fig2}. 
As we see, Eq.~(\ref{ship}) is accurate enough almost everywhere,
for $\Omega\ge 40$,  except in the small vicinity of the obstacle.
Although the obstacle in the theory is represented by a delta function, 
our numerical simulations using a narrow Gaussian potential as the obstacle 
provide results in very good agreement with our extended theory.

\bigskip 
{\bf V. DRAG FORCE} 
\bigskip 

\begin{figure}[ht]
\includegraphics[width=8cm]{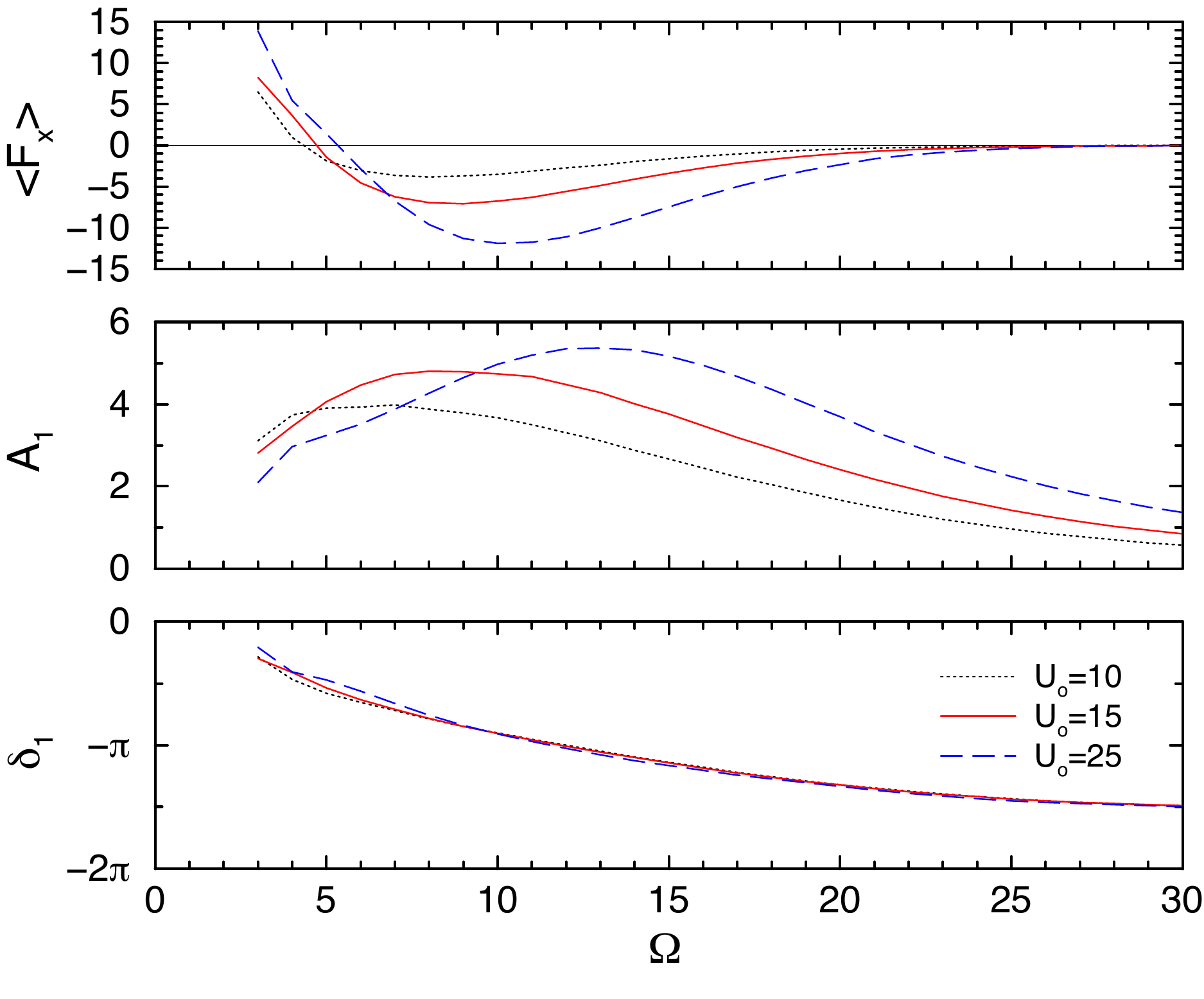}
\caption{Upper panel: Average drag force in the $x$ direction, $<F_x>$, as a function of the 
frequency $\Omega$ for different potential intensities. Middle panel: response function 
amplitude $A_1$ as a function of $\Omega$. Lower panel: relative phase $\delta_1$ as a 
function of $\Omega$. Here we can see that the sign of $<F_x>$ depends on the relative phase 
between the forcing potential and the response main mode, Eq.~(\ref{response}).
}
\label{drag}
\end{figure}

We also computed the drag force in the $x$ direction as
\begin{equation}
F_x(t)=\int_{\mathcal A} \! dxdy \, |\psi|^2 \, \frac{\partial U}{\partial x} \, , 
\end{equation}
where ${\mathcal A}$ defines an infinite region of the fluid around the obstacle. 
For practical purpose we took the integration along our whole grid. 
In Fig.~\ref{drag} we show the average drag taken at one period of oscillation. 
For slow oscillation frequency $\Omega\lesssim5$ we 
observe that drag is decreasing and positive as expected since both non-oscillating attractive and 
repulsive potentials causes positive drag \cite{jacksonattractive}. However, 
for $\Omega\gtrsim5$ the drag is always negative and vanishes in the limit 
of $\Omega\rightarrow\infty$. So, the answer for the question 
{\it can we get rid of drag for very fast oscillations} initially proposed is {\it yes}. 
Surprisingly, the mean drag also vanishes at a small region of low frequencies and this is a 
non-intuitive and remarkable result. 

The system can be seen as a forced oscillator. In our case the oscillating 
potential forces the system and we obtain as output an oscillating $|\psi|^2$. 
Thus the drag force in 
the $x$ direction can be explicitly written as 
\begin{equation}
F_x=U_0\cos(\Omega t)R(t) \, , 
\end{equation}
where $R(t)$ is a response function given by 
\begin{equation}
R(t)\equiv -\frac{4}{w_0^2}  \int_{\mathcal A} x \exp\left[\frac{-2(x^2+y^2)}{w_0^2}\right] |\psi(x,y,t)|^2 dxdy \, .
\end{equation} 
We observed numerically that $R(t)$ is periodic with period $T$ and thus can be written as a Fourier 
series as 
\begin{equation}\label{response}
R(t)=A_0+A_1\cos(\Omega t+\delta_1)+A_2\cos(2\Omega t+\delta_2)+\cdots \, , 
\end{equation}
where $A's$ are amplitudes and $\delta's$ are relative 
phases to the forcing potential. Averaging the drag force in time we have 
\begin{equation}
<F_x>=\frac{1}{T} \int_0^T U_0 \cos(\Omega t) R(t) dt \, ,
\end{equation} 
and only the second term of the series survives giving 
\begin{equation}
<F_x>=\frac{U_0 A_1}{\Omega} \cos(\delta_1) \, .
\end{equation} 
Thus the sign of $<F_x>$ depends on the relative phase between the forcing potential and 
the response main mode. We computed 
\begin{equation}
I\equiv \frac{1}{T}\int_0^TU_0\sin(\Omega t)R(t)dt=-\frac{U_0 A_1}{\Omega} \sin(\delta_1) \, 
\end{equation}
and from $<F_x>$ and $I$ we obtained $A_1$ and $\delta_1$. 

A negative drag can be interpreted as a force in the upwind direction, 
meaning {\it propels} the laser. The question of the energy balance can be explained from the 
oscillating laser. As it attracts and repels the condensate it pumps energy to the system that 
causes an upwind force to supersede the downwind force due to the movement of the laser. This upwind 
force is only generated in the moving and oscillating obstacle. For standing $(M=0)$ oscillating 
obstacle the system is radially symmetric and no drag is generated.

\bigskip
{\bf VI. CONCLUSIONS}
\bigskip

We have studied the wave pattern generated by an oscillating obstacle 
in the supersonic flow of a quantum fluid. Turning on oscillations causes
disruption of the oblique solitons into dipoles. For $\Omega=1.5$ the 
dipoles are emitted organized as a vortex dipole street. For increasing 
frequencies dipoles change gradually orientation in clockwise direction 
and their bunch resembles the oblique solitons. Finally for very high 
frequencies the angle of emission increases and vortices vanishes. For the 
waves in front of the fast oscillating obstacle, we could further extend 
the model previously developed for non-oscillating obstacle. These waves 
were shown to gradually diminish according to the averaging of emission 
of linear waves out of phase. Combined results, both ship waves and soliton 
tend to vanish for high frequencies leading to a vanishing drag. 
Remarkably, the mean drag also vanishes at a small region of low frequencies 
during his change of sign from positive to negative.   
So, even a very powerful laser fast oscillating from red to blue-detuning 
could pass through an atomic BEC without generating vortices or 
solitons. This result could be experimentally checked with existing 
setups \cite{brian}. Analogous experiments could also be performed with 
condensates of exciton-polaritons \cite{amo-2011}.

\bigskip
\begin{center}
{\bf ACKNOWLEDGMENTS}
\end{center}
\bigskip

EGK and AG thanks A.F.R.T. Piza, E.J.V. Passos, E.S. Annibale and F.Kh. 
Abdullaev for useful discussions. EGK thanks the support of G.A. El and the 
hospitality of the Department of Mathematical Sciences, Loughborough, where part of 
this work was carried out. 
We are also grateful to Conselho Nacional de Desenvolvimento Cient\'ifico 
e Tecnol\'ogico and Funda\c{c}\~ao de Amparo \`a Pesquisa do Estado 
de S\~ao Paulo (Brazil) for financial support.

\end{document}